\documentclass[10pt,aps,prd,twocolumn,showpacs,amsmath,amssymb,nofootinbib,eqsecnum,preprintnumbers,superscriptaddress]{revtex4-2}
\usepackage{amsmath,amssymb}
\usepackage{enumitem}  
\usepackage[usenames, dvipsnames]{color} 
\usepackage{graphicx} 
\usepackage{comment}
\usepackage[normalem]{ulem}
\usepackage[utf8]{inputenc}
\usepackage{url}
 \usepackage{xcolor}

\usepackage{hyperref}  

\usepackage{graphicx}
\usepackage{dcolumn}
\usepackage{bm}

\newcommand{\be}{\begin{equation}}
\newcommand{\ee}{\end{equation}}
\newcommand{\ba}{\begin{eqnarray}}
\newcommand{\ea}{\end{eqnarray}}

\newcommand{\beq}{\begin{equation}}
\newcommand{\eeq}{\end{equation}}
\newcommand{\beqa}{\begin{eqnarray}}
\newcommand{\eeqa}{\end{eqnarray}}

\newcommand\feq{\mathrel{\phantom{=}}}

\begin{document}

\preprint{APS/123-QED}

\title{Rotating Carroll Black Holes: A No Go Theorem}%

\author{Ivan Kol{\'a}{\v r}}

\email{ivan.kolar@matfyz.cuni.cz}

\affiliation{Institute of Theoretical Physics, Faculty of Mathematics and Physics,
Charles University, Prague, V Hole{\v s}ovi{\v c}k{\' a}ch 2, 180 00 Prague 8, Czech Republic}

\author{David Kubiz\v n\'ak}

\email{david.kubiznak@matfyz.cuni.cz}

\affiliation{Institute of Theoretical Physics, Faculty of Mathematics and Physics,
Charles University, Prague, V Hole{\v s}ovi{\v c}k{\' a}ch 2, 180 00 Prague 8, Czech Republic}

\author{Poula Tadros}

\email{poula.tadros@matfyz.cuni.cz}

\affiliation{Institute of Theoretical Physics, Faculty of Mathematics and Physics,
Charles University, Prague, V Hole{\v s}ovi{\v c}k{\' a}ch 2, 180 00 Prague 8, Czech Republic}


\date{June 24, 2025}

\begin{abstract}
Recently, there has been a lot of interest in Carroll black holes and in particular whether or not one could find a Carrollian analogue of a rotating black hole spacetime. Here we show that every stationary and axisymmetric solution (and thence also a black hole) of Carrollian general relativity in any number of $d>3$ dimensions
is necessarily also static (up to a ``topological rotation''). The case of $d=3$ dimensions is special. There, the topological rotation is important and one can have a rotating Carroll BTZ black hole, obtained from a static one by the Carroll boost accompanied by the re-identification of the angular coordinate, similar to what happens in the Lorentzian case. 
We also find a Carrollian analogue of an accelerating black hole, showing that Schwarzschild is not the only possible stationary and axisymmetric Carroll 
black hole in four dimensions. A generalization of the no go theorem to include Maxwell, dilatonic, and axionic matter fields is also discussed.
\end{abstract}

 \maketitle

\section{Introduction}
The {\em Carroll algebra}, defined by Levy-Leblond \cite{Levy-Leblond} and independently by Sen Gupta \cite{Gupta}, can be obtained as the In\"on\"u--Wigner contraction of the Lorentz group where the speed of light $c$ goes to zero, ${c\to 0}$, giving rise to an ``ultrarelativistic cousin'' of the Galilean algebra. While the resultant geometry has been predominantly studied from a mathematical point of view, e.g., \cite{Blitz:2025spc}, numerous physical applications have recently begun to emerge.

As of now,
Carroll symmetries are used to describe a wide variety of phenomena, ranging from the memory effect of gravitational waves 
\cite{Gaur:2024oms,Elbistan:2023quw}, to condensed matter physics where fractons are described as Carrollian particles \cite{Figueroa-OFarrill:2023vbj,Bidussi:2021nmp,Ahmadi-Jahmani:2025iqc}. 
Carroll structures also naturally arise
on null surfaces, including black hole horizons 
\cite{Bagchi:2021ban, Donnay:2019jiz} and the asymptotic null infinity \cite{Bagchi:2023cen,Poulias:2025eck,Bagchi:2022emh,Donnay:2022aba,Donnay:2022wvx}. Namely, the fact that 
the asymptotic symmetry group of 
asymptotically flat Lorentzian spacetimes at null infinity,
the Bondi--Metzner--Sachs ($\text{BMS}_4$) group, is isomorphic to the conformal Carroll group in three dimensions motivated the applications to {\em flat space holography}, 
where the asymptotically flat gravitational solutions are conjectured to be dual to 
a conformal Carrollian field theory living on the boundary. Carrollian structures have also been used to describe large scale cosmology
\cite{deBoer:2021jej,Bonga:2020fhx}, 
provide a consistent description of tensionless strings \cite{Bagchi:2019cay},  
capture the behavior of ultrarelativistic fluids
\cite{Freidel:2022bai,Kolekar:2024cfg,Armas:2023dcz, Bagchi:2023ysc, Bagchi:2023rwd}, or to advance our understanding of gravity focusing on generic null hypersurfaces
\cite{Ciambelli:2023mir}.

A {\em Carroll manifold} is a ${d}$-dimensional manifold endowed with a nowhere vanishing vector field ${v}$ and a degenerate metric ${h}$ such that $v^\mu{h_{\mu\nu} = 0}$, where we use  ${\mu, \nu, \ldots \in \{1, \ldots, d\}}$. As mentioned above, this structure can be obtained from a Lorentzian structure by taking the $c \to 0$ limit of the metric, which effectively contracts the light cone at each point to a line. This restricts the mobility of free particles to a single point in space, thus getting the notion of an {\em absolute space}.\footnote{One way to give Carrollian particles non-trivial dynamics is via Hall-like effects, e.g. \cite{Marsot:2022imf, Gray:2022svz}, another is to consider zero energy Carroll particles coming from the limit of relativistic tachyons, e.g. \cite{deBoer:2021jej}.} 
In such manifolds, where space is absolute (and time relative), each point is causally disconnected even from arbitrarily close points, providing a natural foliation of the Carroll manifold by timelike curves (in contrast with Galilean manifolds whose natural foliation is by spacelike hypersurfaces). 

The dynamics of such manifolds is naturally governed by the Carrollian limit of general relativity. The corresponding so called {\em Carrollian General Relativity (CGR)} can be classified into two categories \cite{Henneaux:2021yzg}: {\em electric} \cite{Henneaux:1979vn} and {\em magnetic} \cite{Campoleoni:2022ebj}, derived 
from the Lorentzian general relativity as the leading order and the (particular truncation of) next-to-leading order expansion in $c$, respectively. The resultant theories found their applications to modeling
the deep  {\em strong gravity} regime \cite{Isham:1975ur,Anderson:2002zn,Henneaux:1981su} (see also \cite{Niedermaier:2014ywa} where the same structure was obtained in the limit where the Newton's constant goes to infinity), or to describe the behavior of spacetime 
near singularities (where ultralocality naturally emerges)
\cite{Oling:2024vmq}.

Among the CGR solutions, perhaps the most intriguing are the 
{\em Carroll black holes.} These were defined in \cite{Perez:2022jpr} and \cite{Ecker:2023uwm} as massive solutions to magnetic CGR admitting 
the {\em Carroll extremal surface (CES)},  a locus of points where the Carroll structure breaks down. Such a surface constitutes a structural singularity that cannot be removed by a coordinate transformation, unlike the traditional black hole horizon, which is a mere coordinate singularity and leaves the manifold's differential structure intact. Since Carroll black holes are massive solutions, they cannot exist in the electric theory 
\cite{Perez:2022jpr,Perez:2021abf}, where there is no nonzero charge associated with the timelike Killing vector; all solutions to the electric theory must therefore have zero mass. The notion of Carroll black holes is especially interesting, as such objects feature
non-trivial thermodynamic properties inherited from their Lorentzian cousins
\cite{Ecker:2023uwm,deBoer:2023fnj,Aggarwal:2024yxy}, as well provide novel backgrounds for studying characteristics  of Carroll manifolds, such as geodesics 
\cite{Ciambelli:2023tzb,Chen:2024how}. 

Nevertheless, all $d>3$ Carroll black holes considered in the literature so far are {\em non-rotating}, while all the attempts 
to define a Carroll version of Kerr black hole led to inconsistencies.\footnote{The situation is different in lower dimensions, where a 2D example of a black hole with remnants of rotation was presented in \cite{Ecker:2023uwm}, and a rotating Carroll BTZ black hole was recently constructed in \cite{Aviles:2025ygw}, see  Appendix~\ref{AppA} for further details.}
At the same time,  
it is believed that the Carroll version of a Lorentzian black hole depends on the existence of foliations by spacelike surfaces with vanishing Ricci scalar. Thus, the failure to obtain
a rotating Carroll black hole might have been due to not finding the right coordinates for the Lorentzian solution characterized by such a foliation \cite{Ecker:2023uwm}.

In this paper, we present a general {\em no-go theorem} stating that in all $d>3$ dimensions no matter what coordinates are used to express the Lorentzian black hole, the constraints for CGR forbid rotating solutions (apart from a topological rotation discussed below). The proof goes as follows. We derive the most general stationary, axially symmetric, degenerate metric $h$ along with a nowhere vanishing vector field $v$, such that they satisfy the definition of a Carrollian construction. Then, imposing the ``{\em Hamiltonian constraint''} for magnetic CGR, namely 
\begin{equation}\label{Kmunu}  
K_{\mu\nu}\equiv -\frac{1}{2}\mathcal{L}_{{v}} h_{\mu\nu}=0\;,
\end{equation} 
we conclude that such a construction is non-rotating with no loss in generality. In fact, we show that the same results remain true when the cosmological constant is included,  
as well as seem to survive 
in the presence of 
matter fields. We start by reviewing the Carrollian construction of CGR, following the exposure in \cite{Hansen:2021fxi}.

\section{Carrollian General Relativity}\label{sec2}

In order to write down the corresponding equations of motion in a ``simple way", one introduces two auxiliary objects: the one-form ${T}$, satisfying ${v^{\mu} T_{\mu}=-1}$, and the pseudo-inverse of the metric, ${h}^{-1}$, defined by ${h^{\mu\nu}T_{\nu}=0}$ and ${h_{\mu\sigma}h^{\sigma\nu} = \delta_{\mu}^{\nu}+ v^{\nu}T_{\mu}}$. The latter can be used to raise indices. Note, however, that neither ${h}^{-1}$ nor ${T}$ are objects of the basic Carrollian structure (the field equations can be, in principle, written using only $h$ and $v$).

To derive CGR, one starts from the Einstein--Hilbert action, ${\cal L}=-R$, and employs the \textit{pre-ultralocal (PUL) parametrization} for the metric and its inverse: $g_{\mu\nu}=-c^2T_\mu T_\nu+h_{\mu\nu}$ and $g^{\mu\nu}=-\frac{1}{c^2}v^\mu v^\nu+h^{\mu\nu}$. In this paper, we allow the PUL quantities such as $h_{\mu\nu}$, to have linear terms in $c$ to accommodate rotating metrics. This was done in \cite{Ergen:2020yop} when deriving the non relativistic counterparts of rotating black holes.
The leading order in $c\to 0$ expansion then yields the {\em electric} CGR, whose action reads: 
\begin{equation}
    S_{\text{el}} = \int d^dx E\mathcal{L}_{\text{el}}\;, 
    \quad  
    \mathcal{L}_{\text{el}}=K^2-K^{\mu\nu}\mathcal{K}_{\mu\nu}\;,
\end{equation}
where ${E =\sqrt{\det(T_{\mu},h_{\mu\nu})}}$  is the Carroll-invariant determinant.
The field equations derived from this action are given by two constraints and the evolution equation, namely:
\begin{equation}\label{EOMel}
    \begin{gathered}
      K_{\mu\nu}K^{\mu\nu}- K^2 = 0\;, \quad
      h^{\nu\rho}\nabla_{\rho}(K_{\alpha\nu}-Kh_{\alpha\nu}) = 0\;,\\
     \pounds_v K_{\mu\nu} = -2K_{\mu}^{\lambda}K_{\lambda\nu} + KK_{\mu\nu}\;,
    \end{gathered}
\end{equation}
Here, $\nabla_{\mu}$ is the covariant derivative preserving the Carroll structure, namely $\nabla_{\mu}v^{\nu} =0= \nabla_{\mu}h_{\nu\rho};$ it 
is associated to the Carroll compatible connection $C$ defined as
\begin{equation}
\begin{aligned}
C^{\rho}_{\mu\nu} &= -v^{\rho}\partial_{(\mu}T_{\nu)}- v^{\rho}T_{(\mu}\pounds_{{v}}T_{\nu)} + \tfrac{1}{2} h^{\rho \lambda}\big[ \partial_{\mu} h_{\nu \lambda} 
\\
&\feq + \partial_{\nu} h_{\lambda \mu}- \partial_{\lambda}h_{\mu \nu}\big]- h^{\rho \lambda}T_{\nu}K_{\mu \lambda}\;,
\end{aligned}
\end{equation} 
where ${T_{\mu\nu} \equiv \tfrac{1}{2}(\partial_{\mu}T_{\nu}-\partial_{\nu}T_{\mu})}$.

The above three equations \eqref{EOMel}
act as constraints for the {\em magnetic CGR} whose Lagrangian is given by
\begin{equation}
    \mathcal{L}_{\text{mag}} = - R[C]\;,
\end{equation}
where $R[C]$ is the Ricci scalar for the Carroll compatible connection. 
It was shown in \cite{Hansen:2021fxi} that for stationary solutions this Lagrangian retains its Carroll boost invariance if and only if the Hamiltonian constraint
\eqref{Kmunu}
is imposed. Notice, moreover,  that this constraint is a solution to the electric theory field equations. Upon imposing \eqref{Kmunu} the field equations for the magnetic theory are given by
\begin{equation}\label{Carroll GR equations}
    \begin{aligned}
        M_\mu &\equiv T_{\mu} R[C] + \tfrac{1}{2}h^{\nu\sigma}h_{\mu}^{\lambda}T_{\nu\lambda}v^{\kappa}T_{\kappa\sigma} - \nabla_{\nu}(h^{\nu\sigma}T_{\mu\sigma})\!\!\!\!\! &= 0\;,\\
        M_{\mu\nu} &\equiv R_{(\mu\nu)}[C] - \tfrac{1}{2}h_{\mu\nu}R[C] + \tfrac{1}{2}\nabla_{\lambda}(h_{\mu\nu}h^{\lambda\gamma}v^{\kappa}T_{\kappa\gamma})\!\!\!\!\! &=0\;,
    \end{aligned}
\end{equation}
where $R_{\mu\nu}[C]$ is the Ricci tensor for the Carroll compatible connection. All vacuum stationary Carroll black hole solutions have to satisfy \eqref{Kmunu} and \eqref{Carroll GR equations}. 

To include the {\em cosmological constant}, one typically proceeds as follows. We consider the expansion of the Lorentzian cosmological Lagrangian:
\be\label{Lambdaexpansion} 
-\frac{1}{2}\mathcal{L}_{\text{cosm}} = \tilde \Lambda= \frac{1}{c^2}\Lambda_{\text{el}} + \Lambda + O(c^2)\;.
\ee 
This yields the following electric and magnetic CGR theories:
\ba
    \mathcal{L}_{\text{el}} &=& K^2-K_{\mu\nu}K^{\mu\nu}-2\Lambda_{\text{el}}\;,\nonumber \ \ \ \ 
        \mathcal{L}_{\text{mag}} = - R[C] - 2\Lambda\;.
\ea
If one considers, as per usual \cite{Hansen:2021fxi}, that $\Lambda_{\text{el}}$ is non-zero, the Hamiltonian constraint \eqref{Kmunu} is replaced by
\be\label{KmunuGen} 
K_{\mu\nu}=-\sqrt{\tfrac{\Lambda_{\text{el}}}{d(d-1)}}h_{\mu\nu}\;,
\ee 
which both: i) solves the electric filed equations and ii) ensures the Carroll boost invariance of the magnetic theory. 

Here, instead, we focus on the possibility that ${\Lambda_{\text{el}}=0}$, keeping $\Lambda$ non-trivial \cite{Tadros:2024bev}. This preserves the constraint \eqref{Kmunu} (solving automatically the electric field equations and ensuring the boost invariance) and results in the following modified magnetic CGR equations with cosmological constant:
\begin{equation}\label{GR mag eqs}
 M_\mu = - \frac{2 \Lambda}{d-1}T_{\mu}\;,\quad 
      M_{\mu\nu}=\frac{2 \Lambda}{d-1}h_{\mu\nu}.
\end{equation}
This construction will prove useful in the derivation of the Carroll BTZ metric below.


\section{4D Carroll constructions}\label{4d}
\subsection{General Carroll construction for GR}

Let us consider a stationary axisymmetric Carrollian structure in four dimensions. The most general tensors $v$, $h$, $T$, $h^{-1}$ that are invariant under stationary axial symmetry in 4D are, in spherical coordinates $(t,r,\theta,\phi)$, given by
\ba
v&=&v^\mu\partial_\mu\,,\quad T=T_\mu dx^\mu\,,\nonumber\\
h &=& h_{tt}dt^2+h_{rr}dr^2 +h_{\theta\theta}d\theta^2 +h_{\phi\phi}d\phi^2 + 2h_{t\phi}dt d\phi\,,\nonumber\\
h^{-1}\!&=& h^{tt}\partial_t^2+h^{rr}\partial_r^2 +h^{\theta\theta}\partial_\theta^2 +h^{\phi\phi}\partial_\phi^2 + 2h^{t\phi}\partial_t\partial_\phi\;,\qquad
\ea 
where all functions depend on $r$ and $\theta$. 

Let us first employ the {\em algebraic constraints} of the Carrollian construction. First,
consider the equation ${v^{\mu}h_{\mu\nu}=0}$. Evaluating it for ${\nu=r}$ gives ${v^r=0}$, and for  ${\nu=\theta}$ gives ${v^{\theta}=0}$. When ${\nu=t}$, we get $ h_{tt} = - \tfrac{v^{\phi}}{v^{t}}h_{t\phi}$, whereas 
$\nu=\phi$ yields
$h_{t\phi} = -\tfrac{v^{\phi}}{v^t}h_{\phi\phi}\;.$
We thus have two cases: i) 
$h_{\phi\phi} = 0$, which implies $h_{tt}=0=h_{t\varphi}$ resulting in an extra degenerate 2-dimensional metric, or 
ii) $h_{\phi\phi} \neq 0$ resulting in a degenerate 3-dimensional metric. From now on, we will only consider the latter case.
Considering next
the requirement ${v^{\mu}T_{\mu} = -1}$, yields 
${T_t = -(\frac{1}{v^t}+T_{\phi}V)}$,  
${V\equiv{v^{\phi}}/{v^t}}$. The final step is to define the pseudo-inverse of $h$ satisfying $T_{\mu}h^{\mu\nu}=0$ and $h^{\mu\sigma}h_{\sigma\nu} = \delta^{\mu}_{\nu} + v^{\mu}T_{\nu}$. The second condition, when setting $\mu=t$ and $\nu=r$ and noting that $h^{tr}=0$, gives $T_r=0$, and similarly when $\mu=t$ and $\nu=\theta$, with the observation that $h^{t\theta}=0$, results in $T_{\theta}=0$, while 
the first condition gives
$h^{t\phi}=-\frac{T_t}{T_\phi} h^{tt}=-\frac{T_\phi}{T_t}h^{\phi\phi}$.
The remaining components of the second condition give
$h^{\phi\phi}h_{\phi\phi}=(1+T_{\phi}Vv^t)^2$ and $h^{rr}h_{rr}=1$.
Putting all together, the most general stationary axisymmetric Carrollian construction is given by
\begin{equation}\label{axially symmetric Carroll structure 4d}
    \begin{aligned}
       v &= v^t(\partial_t + V \partial_{\phi})\,,\ \ \ \ T = -(\tfrac{1}{v^t}+ T_{\phi}V)dt + T_{\phi}d\phi\,,\\
        h &= h_{\phi\phi}(Vdt-d\phi)^2+h_{rr}dr^2 +h_{\theta\theta}d\theta^2 \,,\\
        h^{-1} &=\tfrac{1}{h_{\phi\phi}}\Bigl((1{+}T_\phi V v^t)\partial_\phi+T_\phi v^t \partial_t\Bigr)^2+ \tfrac{1}{h_{rr}}\partial_r^2 + \tfrac{1}{h_{\theta\theta}}\partial_{\theta}^2\,,
    \end{aligned}
\end{equation}
where $v^t$, $V$ , $h_{rr}$, $h_{\theta\theta}$, and $h_{\phi\phi}$ are to be determined from the field equations of the theory in question, and $T_{\phi}$ can be (locally) gauge-fixed.

So far the discussion has been entirely algebraic and gravity theory independent. Let us now impose 
the Hamiltonian constraint \eqref{Kmunu} (required by magnetic CGR). Writing the Lie derivative in terms of partial derivatives, and using the fact that, by symmetry, no function depends on $t$ or $\phi$, the constraint reduces to 
$(\partial_{\mu}v^{\sigma})h_{\sigma\nu} +  (\partial_{\nu}v^{\sigma})h_{\sigma\mu} = 0$.
The components ${\mu=t}$, ${\nu=r}$ and ${\mu=r}$, ${\nu=\phi}$ give the equation
$\partial_{r}\ln{V} = 0,$ while the components ${\mu=t}$, ${\nu=\theta}$ and ${\mu=\theta}$, ${\nu=\phi}$ result in $\partial_{\theta}\ln V = 0$
which leads to $V=\mbox{const.}$
This implies that one can perform the following coordinate transformation:
\be \label{transf}
\phi\to \phi+Vt\;.
\ee 
Together with choosing a gauge where ${T_{\phi} = 0}$ (achievable by Carroll boosts), we thus obtained a {\em static} Carrollian structure:
\begin{equation}\label{static construction}
    \begin{aligned}
      v &= v^t\partial_t\;, \ \ \ \ \ \ \
        h= h_{rr}dr^2 + h_{\theta\theta}d\theta^2 + h_{\phi\phi}d\phi^2,\\
        T &= - \tfrac{1}{v^t}dt\;, \ \ \ \
        h^{-1} = \tfrac{1}{h_{rr}}\partial_r^2 + \tfrac{1}{h_{\theta\theta}}\partial_{\theta}^2 + \tfrac{1}{h_{\phi\phi}} \partial_{\phi}^2\;.
    \end{aligned}
\end{equation}
Thus we have proved that in four dimensions every stationary and axisymmetric solution of CGR has to be static, of the form \eqref{static construction} above. As shown in Appendix~\ref{d dim rotation}, the absence of rotation 
remains true for every solution of magnetic CGR in any number of higher dimensions.

Note also that, had we imposed the Hamiltonian constraint with the cosmological constant \eqref{KmunuGen} instead of the `vacuum' constraint \eqref{Kmunu}, the Carrollian structure would break down. Namely, starting from \eqref{axially symmetric Carroll structure 4d}, the constraint \eqref{KmunuGen} reduces to:
\begin{equation}
(\partial_{\mu}v^{\sigma})h_{\sigma\nu} + (\partial_{\nu}v^{\sigma})h_{\sigma\mu} =- \sqrt{\tfrac{\Lambda_{\text{el}}}{6}}h_{\mu\nu}\,.
\end{equation}
From here, the $\mu=\nu=t$ component gives $h_{tt}=0$, the $\mu=\nu=\phi$ component gives $h_{\phi\phi} = 0$, and the $\mu=t,\nu=\phi$ component gives $h_{t\phi} =0$. Also, since ${v^r=v^{\theta}=0}$ and the fact that $h_{rr}$ and $h_{\theta\theta}$ are the only possible remaining components in our construction, the ${\mu=\nu=r}$ component gives ${h_{rr}=0}$, and the ${\mu=\nu=\theta}$ components gives ${h_{\theta\theta}}=0$, i.e., ${h_{\mu\nu}=0}$. Thus, to obtain non-trivial stationary and axisymmetric solutions of magnetic CGR with a cosmological constant, one has to resort to the vanishing cosmological constant in the electric sector of CGR.

\subsection{Einstein-Maxwell-dilaton-axion theory}\label{sec4}
One may wonder whether the above no go theorem is innate to a vacuum (with cosmological constant) CGR, and what would happen upon including matter sector into considerations. For this reason, here we consider a (sufficiently general) extension of CGR, including Maxwell, dilatonic, and also axionic fields. Specifically, 
we start from the {\em Einstein--Maxwell--Dilaton--Axion (EMDA)} theory, which arises from the
low energy effective action of superstring theories upon compactifying the ten-dimensional
heterotic string theory on a six-dimensional torus \cite{Sen:1992ua}. 

The Lorentzian Lagrangian for this theory is given by \begin{equation}\label{dilaton axion Lagrangian}
\begin{aligned}
    \mathcal{L} &= - R + 2\partial_{\mu}\phi\partial^{\mu}\phi + \tfrac{1}{2}e^{4 \phi}\partial_{\mu}\kappa\partial^{\mu}\kappa\\
    &\feq - e^{-2\phi}F_{\mu\nu}F^{\mu\nu}- \kappa {(*F)}_{\mu\nu}F^{\mu\nu}\;,
    \end{aligned}
\end{equation}
where $R$ is the Ricci scalar of the Levi--Civita connection, $\phi$ is a scalar field, $\kappa$ is the axion and $F_{\mu\nu}$ is the electromagnetic tensor and ${(*F)}_{\mu\nu}$ denotes its Hodge dual. 

To find the Carroll expansion of the above theory, one again employs the PUL parametrization for the metric, together with the following decomposition for the electromagnetic field:
$F_{\mu\nu}F^{\mu\nu} = \tfrac{1}{c^2}(-2 E^2)+B^2$, where $E_{\mu} = v^{\nu}F_{\mu\nu}$ and $B^2 = B_{\mu\nu}B^{\mu\nu}$ with $B_{\mu\nu}=h_{\mu\sigma}h_{\nu \rho}F^{\sigma\rho}$, and ${(*F)}_{\mu\nu}F^{\mu\nu}=\tfrac{2}{c}E_{\mu}B^{\mu} + O(c^{-3})$, where $B^{\mu} = T_{\nu}\epsilon^{\nu\mu\alpha\beta}F_{\alpha\beta}$ 
\cite{deBoer:2021jej, deBoer:2023fnj}. 
The leading order expansion then yields the 
electric theory, as usual. However, the coupling of the axion to the second (parity violating) electromagnetic invariant $(*F)_{\mu\nu}F^{\mu\nu}$, induces a term of order $c^{-1}$, generating yet another Lagrangian, $\mathcal{L}_{\text{mix}}$, between the usual electric and magnetic Lagrangians:\footnote{Note that in the presence of additional matter fields one has a freedom of rescaling  these fields in various ways  in $c$, resulting in a number of possible Carroll theories. In here we have focused on one particular such limit of the EMDA theory. } \begin{equation}\label{(0,0)}
    \begin{aligned}
\mathcal{L}_{\text{el}} &= -K^2+K_{\mu\nu}K^{\mu\nu} + 2 (v^{\sigma}\partial_{\sigma}\phi)^2 +  \tfrac{1}{2}e^{4\phi}(v^{\sigma}\partial_{\sigma}\kappa)^2 \\ &\feq + 2 e^{-2\phi}E^2\;, \quad \mathcal{L}_{\text{mix}} =  -2 \kappa E_{\mu}B^{\mu}\;,
\\
\mathcal{L}_{\text{mag}} &= R[C] \!+\! 2 h^{\mu\nu}\partial_{\mu}\phi\partial_{\nu}\phi \!+\! \tfrac{1}{2} e^{4\phi}h^{\mu\nu}\partial_{\mu}\kappa\partial_{\nu}\kappa  \!-\! e^{-2\phi}B^2\;.
    \end{aligned}
\end{equation}

At the electric level, the variations w.r.t. the Carrollian structures yield the following constraints:
\begin{equation}\label{electric eqns}
    \begin{aligned}
        &\pounds_v K_{\mu\nu} = -2K_{\mu}^{\lambda}K_{\lambda\nu} + KK_{\mu\nu} - 7h_{\mu\nu}(v^{\sigma}\partial_{\sigma}\phi)^2 \\&\qquad\quad + 2e^{4\phi}h_{\mu\nu}(v^{\sigma}\partial_{\sigma}\kappa)^2 +  e^{-2\phi}(4E_{\mu}E_{\nu} - \tfrac{1}{2}h_{\mu\nu}E^2)\;,\\
       &K_{\mu\nu}K^{\mu\nu}-K^2 + 10 (v^{\sigma}\partial_{\sigma}\phi)^2  + \tfrac{5}{2}e^{4\phi}(v^{\sigma}\partial_{\sigma}\kappa)^2 \\& 
       \qquad\qquad + 2e^{-2\phi}E^2 = 0\;,\\
    &h^{\nu\rho}\nabla_{\rho}(K_{\alpha\nu}-Kh_{\alpha\nu}) + 4 (v^{\sigma}\partial_{\sigma}\phi)\partial_{\alpha}\phi \\ &\qquad \qquad+ e^{4\phi}(v^{\sigma}\partial_{\sigma}\kappa)\partial_{\alpha}\kappa  +4  e^{-2\phi}E^{\nu}B_{\nu\alpha} = 0\;.
            \end{aligned}
\end{equation}
These are to be accompanied by the corresponding Maxwell, axion, and dilaton electric equations of motion, and by the equations coming from the mixed Lagrangian, in the case of non-trivial axion (see App.~\ref{App:C} for more details).

The general construction \eqref{axially symmetric Carroll structure 4d} is now supplemented by the dilaton and axion fields, which we assume to be functions of $r$ and $\theta$ only. The dilatonic field equations then imply that that the electric field has to vanish, $E_\mu=0$. On the other hand, when the dilaton is absent in the theory (for example when we consider the pure Einstein--Maxwel case), 
the constraint $E_{\mu}=0$ cannot be directly inferred from the field equations. To proceed further
we simply have to assume that also in this case, and since we are interested in magnetic solutions, one can set
\be 
E_\mu=0\,. 
\ee
(Note that this assumption is typically satisfied when one obtains the Carrollian solution from a Lorentzian one, e.g. \cite{deBoer:2023fnj}.)

With these assumptions, it is then easy to verify that the electric constraints \eqref{electric eqns} now reduce to the vacuum CGR equations of motion \eqref{EOMel}, while the remaining electric (matter field) equations are trivially satisfied. Thus, again one  needs to impose the Hamiltonian constraint \eqref{Kmunu} and any solution of the corresponding Carrollian EMDA theory has to be static, as in the CGR vacuum case.

\section{3D Carroll construction}\label{3d}

The case of three dimensions is special. Namely, similar to the Lorentzian case, any solution of magnetic CGR with cosmological constant is locally the Caroll version of (A)dS$_3$. To show this, we first note that the torsion vanishes as a consequence of (2.15) in \cite{Hansen:2021fxi} when the constraint \eqref{KmunuGen} is used with ${\Lambda_{\text{el}} = 0}$. Furthermore, taking the trace of the second  magnetic equation in \eqref{GR mag eqs}, we find that ${\nabla_{\lambda}(h^{\lambda\gamma}v^{\kappa}T_{\kappa\gamma}) = 2\Lambda}$. Inserting this back, we observe that the traceless Ricci tensor vanishes, ${R_{(\mu\nu)}[C] - \tfrac{1}{2}h_{\mu\nu}R[C] = 0}$, while inserting it into the first magnetic equation in \eqref{GR mag eqs}, contracted with $v^{\mu}$, implies that the Ricci scalar is constant, ${R[C] = -3\Lambda}$. Recalling that the Weyl tensor is identically zero in 3D, the Riemann tensor decomposition then yields
\begin{equation}
R_{\mu\nu\sigma\rho}[C] = \tfrac{-\Lambda}{2}(h_{\mu\sigma}h_{\nu\rho} - h_{\mu\rho}h_{\nu\sigma})\;,
\end{equation}
which implies that the spacetime is of constant curvature, ${\nabla_{\kappa} R_{\mu\nu\sigma\rho}[C] = 0}$.
Moreover, since $v^\mu$ has to satisfy the magnetic equations \eqref{GR mag eqs}, one can show that locally both structures $h_{\mu\nu}$ and $v^\mu$ match those of Carroll (A)dS$_3$.



However, this does not mean that black holes cannot exist in $(2+1)$-dimensional CGR. In fact, repeating the 
above local construction, starting from the general elements in polar coordinates $(t,r,\phi)$, and upon imposing the algebraic and Hamiltonian constraints (together with the transformation \eqref{transf}), we arrive at:
\ba \label{axially symmetric Carroll structure}
v &=& v^t \partial_t\,,\ \, 
h = h_{\phi\phi}d\phi^2+h_{rr}dr^2\,,\ \, T = -\tfrac{1}{v^t}dt + T_{\phi}d\phi\,,\nonumber\\
&&h^{-1} =\tfrac{1}{h_{\phi\phi}}\big(\partial_\phi+T_\phi v^t \partial_t\big)^2+ \tfrac{1}{h_{rr}}\partial_r^2\;,
\ea
where $v^t$, $V$, $h_{rr}$, and $h_{\phi\phi}$ are functions of $r$, to be determined from the field equations of the theory in question, and $T_{\phi}$ can be (locally) gauge-fixed (see below).

In particular, setting $T_\phi=0$ (achievable by Carroll boost), 
we can now impose the magnetic CGR equations of motion with cosmological constant, \eqref{GR mag eqs}, parameterizing further $\Lambda=-1/\ell^2$. Without loss of generality, we can set i) $h_{\phi\phi}=1$ or ii) $h_{\phi\phi}=r^2$. One can easily show that the former leads to inconsistencies in the field equations and we focus on the latter. Denoting further $f=1/h_{rr}$, and after some algebra, the $rr$ component of the second equation in \eqref{GR mag eqs},  reduces to $f'=2r/\ell^2$. Moreover, the 
first equation gives $v^t=1/\sqrt{f}$, with the remaining equations satisfied. Thus, we have re-obtained the {\em static Carroll BTZ black hole} \cite{Aviles:2024llx}:
\be 
v=\frac{1}{\sqrt{f}}\partial_t\,,\quad h=\frac{dr^2}{f}+r^2d\phi^2\,,\quad f=\frac{r^2}{\ell^2}-m\,,
\ee 
where $m$ is an integration constant ($m>0)$.
The corresponding CES is located at $f(r_{\mbox{\tiny CES}})=0$, that is, $r_{\mbox{\tiny CES}}=\ell\sqrt{m}$. This black hole features thermodynamic properties inheritted from the Lorentzian BTZ black hole. In particular, it can be assigned temperature and entropy obeying the first law  (see Appendix~\ref{AppA} for details and 
further discussion of properties of this black hole).

The {\em rotating Carroll BTZ} can then, similar to its Lorentzian counterpart \cite{Clement:1995zt, Martinez:1999qi}, be obtained by applying the Carrollian boost
accompanied by re-identification of the angular coordinate. Such a boost, which only affects $T$ and $h^{-1}$ (while the Carrollian structures $v$ and $h$ remain invariant) amounts to re-introducing  $T_\phi=J\sqrt{f}$ in \eqref{axially symmetric Carroll structure}. In \cite{Aviles:2025ygw}, this was shown to introduce the topological (global) angular momentum $J$ into the  
spacetime, 
giving rise to the rotating Carroll BTZ for the first time. 
Let us stress that while locally $T_\phi$ can be set to zero, it is the global re-identification of the angular coordinate that introduces this rotation in 3D. Of course a similar construction also works in higher dimensions, where, however, it is not very interesting as it simply amounts to a global threading of spacetime with a rotating string.

\section{Conclusions}\label{sec5}

In this note we have proved the following:
{\bf Theorem:} {\em  Any stationary and axially symmetric solution of magnetic CGR in any number of $d>3$ dimensions is necessarily static.}
In particular, this implies that there are no axisymmetric and rotating Carrollian black holes.   
Thus, the failure to obtain  
one in the literature is not due to not finding the right coordinate system to carry out the Carrollian limit,  
but rather to the features of CGR. 

In 3D, the situation is different and a rotating Carroll black hole can be obtained by introducing the topological rotation via identifications, similar to what happens in the Lorentzian case.  
At the same time, 
one can easily show that the Carrollian Schwarzschild is not the only stationary and axisymmetric Carroll black hole in 4D. 
Namely, one can easily give an example of a Carroll black hole that is not spherically symmetric, namely, the Carroll version of the C-metric (e.g. \cite{Griffiths:2006tk}). Since the Lorentzian C-metric is static, one can straightforwardly write its Carroll version, which reads:
\ba
h &=& \frac{1}{(1+\alpha r \cos\theta)^2}\Big(\frac{dr^2}{Q} + \frac{r^2d\theta^2}{P}+ P r^2\sin^2\!\theta d\phi^2\Big)\,,\nonumber\\
v &=& - \frac{1+\alpha r \cos\theta}{\sqrt{Q}}\partial_t\,,
\ea
where $P = 1+2\alpha m \cos\theta$, ${Q = (1-\alpha^2 r^2)(1-{2m}/{r})}$, and parameters $\alpha$ and $m$ are the acceleration, mass parameters, respectively. 
A generalization of this metric metric to AdS is rather interesting from a thermodynamic point of view and will be discussed elsewhere.

Finally, we have shown that (modulo some caveats) the above no go theorem 
extends to the 4D EMDA theory -- the addition of matter does not seem to change the conclusion that stationary axially symmetric Carroll black holes in the Carrollian version of this theory must be static --- it only further restricts the matter fields.
It remains to be seen whether this conclusion is generic and remains true for an arbitrary matter content and in any number of dimensions.

\subsection*{Acknowledgments}
We would like to thank Oscar Fuentealba, Finnian Gray, Daniel Grumiller, Marc Henneaux, Robie Hennigar, Niels Obers, and Watse Sybesma 
for discussions and reading the manuscript. I.K. and P.T. 
acknowledge financial support by Primus grant PRIMUS/23/SCI/005 from Charles University.
D.K. is grateful for support from GA\v{C}R 23-07457S grant of the Czech Science Foundation. D.K. and I.K. also received support from the Charles University Research Center Grant UNCE24/SCI/016. P.T. acknowledges financial support by GAUK 425425 from Charles University.

\appendix 

\section{Carroll BTZ black hole}\label{AppA}

\subsection{BTZ and its Carrollian limit}
The Lorentzian BTZ metric is given by \cite{Banados:1992wn} 
\begin{equation}\label{BTZ metric}
    ds^2 = -c^2f dt^2 + \frac{dr^2}{f} + r^2(d\phi+ N^{\phi} cdt)^2,
\end{equation}
where 
\be
f=-\frac{8 G_3 M}{c^2} + \frac{r^2}{\ell^2} + \frac{16 G_3^2 J^2 }{c^6 r^2}\,,\quad N^{\phi} = -\frac{4G_3J}{c^3 r^2}\,,
\ee
recovering $G_3$'s and $c$'s in the expressions. The spacetime features a black hole horizon, given by the largest root of $f(r_h)=0$; it is characterized 
by the asymptotic charges $M$ and $J$, and features the following thermodynamic quantities: 
\ba \label{BTZ_TDs_Lor}
M&=&\frac{c^2 r_h^2}{8G_3 \ell^2}+\frac{2G_3 J^2}{c^4 r_h^2}\,,\nonumber\\
T&=&\frac{c f'(r_h)}{4\pi}=\frac{
c r_h}{2\pi \ell^2}-\frac{8 
G^2_3 J^2}{\pi c^5 r_h^3}\,,\\
S&=&\frac{\pi c^3 r_h}{2G_3 
}\,,\quad \Omega=\frac{4 G_3J}{c^2 r_h^2}\,. \nonumber
\ea 
These obey the first law of black hole thermodynamics 
\be 
c^2 \delta M=T\delta S+\Omega \delta J\,,
\ee 
One may also include the variations of the cosmological constant, adding a $P-V$ term into black hole thermodynamics, e.g. \cite{Kubiznak:2016qmn}. Namely, one has 
\be 
P=\frac{c^4}{G_3}\frac{\Lambda}{8\pi}=\frac{c^4}{8\pi G_3\ell^2}\,,\quad V=\pi r_h^2\,,
\ee
upon which the following extended first law works:
\be 
c^2 \delta M=T\delta S+\Omega \delta J+V\delta P\,.
\ee 
This is then accompanied by the following Smarr relation:
\be 
0=TS+\Omega J-2PV\,,
\ee 
which follows from the Euler scaling argument and dimensional analysis. On the other hand, by keeping the horizon radius fixed while we scale:
$J\to \alpha J, \ell\to \ell/\alpha$ yields:
\be 
c^2 M=\frac{1}{2}\Omega J+PV\,.
\ee 
Of course, one can also combine the two to get
\be
c^2M=\Omega J+\frac{1}{2}TS\,,
\ee
which is yet another Smarr relation valid for BTZ black holes.

To take the Carroll limit, 
we expand the cosmological term according to \eqref{Lambdaexpansion}, setting $\Lambda_{\mbox{\tiny el}}=0$ and keeping $\Lambda$ non-trivial, as discussed in the main text. This amounts to `preserving' the AdS length scale $\ell\to \ell$. At the same time, we perform the following scaling:
\be \label{rescalings}
G_3 = c^4 \bar{G}_3\,,\quad  M = \frac{\bar{M}}{c^2}\,,\quad  
J = \frac{\bar{J}}{c}\,,
\ee 
to have a consistent Carroll limit with consistent thermodynamics \cite{Ecker:2023uwm, Ecker:2024czh}.
Substituting and rearranging the metric, we get 
\be \label{fbar}
\bar{f} = -8 \bar{G}_3 \bar{M} + \frac{r^2}{\ell^2}+ 
\frac{16 \bar{J}^2 \bar{G}_3^2}{r^2}\,,\quad  
\bar{N^{\phi} } = - \frac{4 \bar{G}_3 \bar{J}}{r^2}\,.
\ee

To cast the metric in the PUL form, we solve the parameterization equations $g_{\mu\nu} = -c^2 T_{\mu}T_{\nu}+h_{\mu\nu}$ and $g^{\mu\nu} = -\tfrac{1}{c^2}v^{\mu}v^{\nu}+h^{\mu\nu}$, where $g_{\mu\nu}$ are the components of the Lorentzian metric and $T_{\mu}, v^{\mu}, h^{\mu\nu}$ and $h_{\mu\nu}$ are of the form  \eqref{axially symmetric Carroll structure}. Notice that these quantities satisfy $T_{\mu}v^{\mu}=-1$, $v^{\mu}h_{\mu\nu}=0, T_{\mu}h^{\mu\nu}=0$ and $h_{\mu\sigma}h^{\nu\sigma} = \delta_{\mu}^{\nu}+v^{\nu}T_{\nu}$. Solving these equations we get $v^{t}=\tfrac{1}{\sqrt{\bar{f}+r^2(\bar{N}^{\phi})^2}}$, $h_{rr}=\tfrac{1}{\bar{f}}$, $h_{\phi\phi}= r^2$, $v^{\phi}= - 2c \tfrac{\bar{N}^{\phi}}{\sqrt{\bar{f}}}$ and $V = c\bar{N}^{\phi}$. This construction is Carrollian, however, it is not necessarily a solution for CGR unless we impose the constraint $V=$ constant derived in the main text. To satisfy this constraint, we have to set $\bar{N}^{\phi}=$ constant, which is not true except at 
\be \label{J0}
\bar{J}=0\,. 
\ee 
So, we end up with
\be\label{carroll btz}
     v = \frac{1}{\sqrt{\hat f}} \partial_t\,, \quad h = \frac{dr^2}{\hat f}+r^2d\phi^2\,,\quad \hat f= -8 \bar{G}_3 \bar{M} + \frac{r^2}{\ell^2}\,,
    \ee
which is the solution derived in the main text. It also coincides with the Carroll limit of the static BTZ black hole. This solution can further be endowed with rotation via Carrollian boost accompanied by re-identification of the angular coordinate, as discussed in the main text.

Note that there is yet another extra degenerate case where $h_{\phi\phi} = 0$ in the general Carroll construction \eqref{axially symmetric Carroll structure}. Comparing  \eqref{BTZ metric} with the extra degenerate Carroll metric, it is evident that it 
cannot be written as a PUL parameterization of such metric except if the terms $r^2(d\phi + N^{\phi}cdt)^2$ were removed,
leading to the following geometry:
\be 
v = \frac{1}{\sqrt{\bar f}}\partial_t\,,\quad 
        h = \frac{dr^2}{\bar f}\,,
\ee
where $\bar f$ is given in \eqref{fbar}.
Note that in this case, we can preserve non-trivial $\bar J$.
The same geometry was obtained in \cite{Ecker:2023uwm} by the Kaluza--Klein reduction of the BTZ geometry to charged 2D dilaton gravity, followed by the Carrollian limit.

\subsection{Thermodynamics}

The Carroll BTZ black hole \eqref{carroll btz} inherits thermodynamic properties from its Lorentzian counterpart. 
First, it admits the CES. It happens where the Carroll vector $v$ degenerates, that is, at 
\be \label{CEScon}
{\hat f}(r_{\mbox{\tiny CES}})=0\,.
\ee 
Formally, this CES can be assigned the following entropy and temperature:
\ba
S &=& \frac{\mbox{Area(CES)}}{4\bar{G}_3c} = \frac{\pi r_{\text{\tiny CES}}}{2\bar{G}_3c}\,,\nonumber\\
T&=&\frac{c{\hat f}'(r_{\mbox{\tiny CES}})}{4\pi}=\frac{cr_{\mbox{\tiny CES}}}{2\pi \ell^2}\,.
\ea
It is easy to verify that these quantities can be also straightforwardly obtained from the Lorentzian ones \eqref{BTZ_TDs_Lor} upon taking into account the rescalings \eqref{rescalings}, together with the fact that we have to set $\bar J=0\,,$ \eqref{J0}, and replace $r_h\to r_{\mbox{\tiny CES}}$. Moreover, such a limit yields the following expressions for the Carroll mass and pressure/volume terms: 
\be 
\bar M=\frac{r^2_{\mbox{\tiny CES}}}{8 {\bar G}_3}\,,\quad 
P=\frac{1}{8\pi \bar G_3 \ell^2}\,,\quad V=\pi r_{\mbox{\tiny CES}}^2\,,
\ee 
whereas the expressions for $\Omega$ and $\bar J$ vanish. 
Note that the expression for $\bar M$ coincides with the one obtained from the CES condition \eqref{CEScon}. It is now easy to verify that the following first law holds: 
\be 
\delta \bar M=T\delta S+V\delta P\,,
\ee
and it is accompanied by these Smarr relations:
\be 
TS=2P V\,,\quad 
{\bar M}= PV\,,\quad {\bar M}=\frac{1}{2}TS\,.
\ee 
Notice, however, that similar to other Carroll black holes, the entropy diverges in the strict limit $(c\to 0)$ while the temperature goes to zero, leaving the energy finite, e.g. \cite{Ecker:2023uwm}.

\subsection{Geodesics}

Similar to the Lorentzian case, geodesics can uncover properties of Carroll manifolds. Following 
\cite{Ciambelli:2023tzb}, we write the action for a particle in the Carroll geometry as:
\begin{equation}
    I = \int d\tau e \big(g_0 + \frac{g_1}{e}h_{\mu\nu}\dot{x}^{\mu}\dot{x}^{\nu} + \frac{g_2}{e^2}\dot{t}^2\big)\;,
\end{equation}
where $e$ is an auxiliary field, and $g_0, g_1$, and $g_2$ are constants. We stress that such an action describes the motion of `normal' particles in Carrollian geometry, rather than the motion of Carroll particles (which actually cannot move).

Using the static Carroll BTZ derived above, \eqref{carroll btz},
we get
\begin{equation}
     I = \int d\tau e \Big[g_0 + \frac{g_1}{e}\Big(\frac{\dot{r}^2}{\hat{f}}+r^2 \dot{\phi}^2\Big)+ \frac{g_2}{e^2}\dot{t}^2\Big]\;.
\end{equation}
The equations of motion for $t$ and $\phi$ are straightforward:
\begin{equation}
    \begin{aligned}
        t = c_1 \tau + c_2, \ \ \ \ \dot{\phi} = \frac{L}{r^2}\;,
    \end{aligned}
\end{equation}
where $c_1$, $c_2$, and $L$ are integration constants. The radial equation can be derived from the equation of motion for the auxiliary field $e$ by parameterizing ${e=\text{const.}}$:
\begin{equation}
    \sqrt{\frac{g_1}{g_0}h_{\mu\nu}\dot{x}^{\mu}\dot{x}^{\nu} + \frac{g_2}{g_0}\dot{t}^2} = e = \text{const.}
\end{equation}
Defining ${{E}/{\bar{M}} \equiv ({g_0}/{2g_1})e^2-({g_2}/{2g_1})c_1^2}$, 
we get the radial geodesic equation
\begin{equation}
    \frac{\dot{r}^2}{2} + V_{\text{eff}}(r) = -4 \bar{G}_3 E-\frac{L^2}{2\ell^2}\;,
\end{equation}
where ${V_{\text{eff}}(r) \equiv -{2 r^2 E}/{(\bar{M}\ell^2)} - {\bar{M} \bar{G}_3 L^2}/{(2r^2)}}$ is the effective potential.

To derive the circular orbits, we impose the circularity conditions: i) ${\dot{r}=0}$  (i.e., ${V_{\text{eff}}(r) = -{2 r^2 E}/{(\bar{M}\ell^2)} - {\bar{M} \bar{G}_3 L^2}/{(2r^2)}}$), ii) ${\ddot{r}=0}$ (i.e., ${V'_{\text{eff}}(r)=0}$). A convenient way to do so is to write the radial equation as
\begin{equation}
    \dot{r}^2 = \Bigl(-8 \bar{G}_3 \bar{M} + \frac{r^2}{\ell^2}\Bigr)\Bigl(\frac{1}{b^2}-\frac{1}{r^2}\Bigr)\frac{L^2}{2}\;,
\end{equation}
where ${b = {\sqrt{\bar{M}} L}/{\sqrt{E}}}$. Thus, (i) implies either ${r=r_{\text{\tiny CES}}}$, or ${r=b}$ and (ii) gives ${r^2 = {\bar{M} L \sqrt{G}_3 \ell}/{2 \sqrt{E}}}$. If ${r=r_{\text{\tiny CES}}}$, we get also ${b=r_{\text{\tiny CES}}}$. The second derivative of the effective potential,
\begin{equation}
    V''_{\text{eff}}(r_{\text{CES}}) = -\frac{1}{\bar{M} \ell^2}\Bigl(4E + \frac{3 L^2}{64 \bar{G}_3\ell^2}\Bigr)<0\;,
\end{equation}
then tell us that the orbit is unstable. The case ${r=b}$ with ${r\neq r_{\text{\tiny CES}}}$ contradicts (ii). Thus, the only circular orbit is on CES and it is unstable. This result is similar to those of Carrollian  Schwarzschild, Schwarzschild-(A)dS, Schwarzschild--Bach, Schwarzschild--Bach-(A)dS, and Reissner--Nordstr\"om  \cite{Ciambelli:2023tzb,Chen:2024how,Tadros:2024bev}.

\section{Carroll construction in a general dimension} \label{d dim rotation}

In this appendix we show that stationary and axisymmetric Carroll black holes in any number of higher dimensions cannot rotate. In what follows we adopt the notation employed in  \cite{Myers:2011yc}.

In $d$ number of dimensions 
the maximum number of rotation parameters is $n =[{(d-1)}/{2}]$ (the integer part of ${(d-1)}/{2}$). The corresponding Carroll construction is 
\begin{equation}
    \begin{aligned}
    v&= v^t\partial_t + v^r\partial_r + 
\epsilon v^{\alpha}\partial_{\alpha} + \sum_{i=1}^n\bigl({\bar v}^{i}\partial_{\mu_i}+ v^i\partial_{\phi_i}\bigr)\;,\\
    h &= h_{tt}dt^2 + h_{rr}dr^2 + \epsilon h_{\alpha\alpha}d\alpha^2 +   \sum_{i=1}^n h_{ti}d\phi_idt \\
    &\feq +\sum_{i,j=1}^n \bigl({\bar h}_{ij}d\mu_id\mu_j+ h_{ij}d\phi_id\phi_j\bigr)\;,
    \end{aligned}
\end{equation}

with the constraint
\begin{equation}
\sum_{i=1}^n\mu_i^2+\epsilon \alpha^2=1\;,
\end{equation}
where ${\epsilon = 1}$ for even $d$ and ${\epsilon = 0}$ for odd $d$. Here, to reflect stationarity and axisymmetry, all metric functions and components of $v$ are assumed to be functions of $r$ and $\mu_i$ coordinates only (while coordinate $\alpha$ is eliminated via the constraint).

To show that the vacuum magnetic CGR implies staticity, we proceed as in the main text. First, we impose the algebraic constraints. 
The condition ${v^{\mu}h_{\mu\nu} = 0}$ in odd $d$ translates into
\begin{equation}\label{d dim orth cond}
    \begin{aligned}
    & v^{r} =0= {\bar v}^{k}\;, 
    \\
       &v^th_{tt} + \sum_{i=1}^nv^i h_{t i} = 0\;,\\
       & v^t h_{t k} + \sum_{i=1}^nv^i h_{k i} = 0\;,
    \end{aligned}
\end{equation}
for all $k \in \{1,\ldots,n\}$. 
For even $d$, we have to add the constraint ${v^{\alpha}=0}$. The non-zero components of the CGR constraint ${K_{\mu\nu} = 0}$ are the $ri$ components, which lead~to
\begin{equation}
    \begin{aligned}
        \partial_r v^t h_{k t} + \sum_{j=1}^n \partial_r v^j h_{kj} = 0\;,
    \end{aligned}
\end{equation}
for all ${k \in \{1,\ldots,n\}}$. Using \eqref{d dim orth cond}, we get
\begin{equation}
    \sum_{j=1}^nh_{ij} v^j \partial_r \ln\Big(\frac{v^j}{v^t}\Big) = 0\;,
\end{equation}
for all ${i \in \{1,\ldots,n\}}$. From this constraint we see that in the case of one axis of rotation ($n=1$), we get ${\partial_r \ln({v^i}/{v^t}) = 0}$. Same can be said about the $\mu$ dependencies of $\ln(v^i/v^t)$ by considering the $r\mu_i$ components of $K_{\mu\nu}=0$,  which means ${{v^i}/{v^t} =\text{const.}}$ Therefore, we can transform the Carroll construction into a static one 
as in the 4D and 3D cases. 

This is also possible for the case with $n$ symmetry axes of rotation in arbitrary number of dimensions $d$. For brevity, however, we will only demonstrate it explicitly on the 5D example with two independent rotations. Then, the constraints on the Carroll construction read
\begin{equation}\label{5d constraints}
    \begin{aligned}
        &v^t h_{tt} = - v^1 h_{t1} - v^2 h_{t 2}\;,\\
        & v^t h_{i t} = - v^1 h_{i1} - v^2 h_{i2}\;,
    \end{aligned}
\end{equation}
for ${i=1,2}$. From the first two equations, we have
\begin{equation}
    h_{tt} = \Big(\frac{1}{v^t}\Big)^2\Big[(v^1)^2 h_{11} + v^1v^2h_{12} + (v^2)^2h_{22}\Big]\,.
\end{equation}
We thus have two cases: i) $(v^1)^2 h_{11} + v^1v^2h_{12} + (v^2)^2h_{22}=0$, which implies that $h_{tt}=0$, thus, resulting in an extra degenerate metric, so we will not explore this case further. ii) $(v^1)^2 h_{11} + v^1v^2h_{12} + (v^2)^2h_{22} \neq 0$. In this case the $tr$, $r1$, $r2$ components of the CGR constraint read
\begin{equation}\label{constraints in 5d}
    \begin{aligned}
        &\frac{\partial_rv^t}{v^t}\big[(v^1)^2 h_{11} + v^1v^2h_{12} + (v^2)^2h_{22}\big] \\
        &\ - \partial_rv^1(v^1h_{11}+v^2h_{12}) - \partial_rv^2(v^1h_{12}+v^2h_{22})=0\,,\\
        &-\frac{\partial_rv^t}{v^t}(v^1h_{11}+v^2h_{12}) + (\partial_rv^1)h_{11} + (\partial_rv^2)h_{12}=0\,,\\
        &-\frac{\partial_rv^t}{v^t}(v^1h_{12}+v^2h_{22}) + (\partial_rv^1)h_{12} + (\partial_rv^2)h_{22}=0\,.
    \end{aligned}
\end{equation}
multiplying $v^1$ by the second equation, and $v^2$ by the second, and substituting in the first, we get
\begin{equation}
    v^1v^2h_{12}\partial_rv^t=0\,.
\end{equation}
From the construction, we assumed that $v^1$ and $v^2$ are non-zero, otherwise it would be a black hole with a single rotation which we considered before. We will also assume that $\partial_rv^t \neq 0$ since if it is the case, we can use the $r\mu_i$ components of the constraints and deduce that $v^t$ is a constant which is not consistent. Thus, we have to set $h_{12}=0$. Doing so and using the second and third equations in \eqref{constraints in 5d}, we get
\begin{equation}
    \partial_r\ln(v^1/v^t) = \partial_r\ln(v^2/v^t) = 0\,.
\end{equation}
The dependencies on $\mu_i$ can be derived in the same was from the $r\mu_i$ components of the CGR constraint. Thus, the metric can be written in a static form in the same way as the 3D and 4D cases.

\section{Carrollian EMDA theory}\label{App:C}

\subsection{Useful identities}
When deriving the Carrollian limit of the EMDA theory, one needs to re-express the two electromagnetic invariants. Let us gather here the corresponding identities.

First, we have 
$F_{\mu\nu}F^{\mu\nu} = g^{\mu\nu}g^{\sigma\rho}F_{\mu\sigma}F_{\nu\rho}.  
$
Upon using the PUL form \cite{Hansen:2021fxi}: ${g^{\mu\nu} = -\tfrac{1}{c^2}v^{\mu}v^{\nu}+h^{\mu\nu}}$, we get the following non-zero terms:
\begin{equation}
\begin{aligned}
    F_{\mu\nu}F^{\mu\nu} &= -\tfrac{1}{c^2}(v^{\mu}v^{\nu}h^{\sigma\rho} + v^{\sigma}v^{\rho}h^{\mu\nu})F_{\mu\sigma}F_{\nu\rho} \\&+ h^{\mu\nu}h^{\sigma\rho}F_{\mu\sigma}F_{\nu\rho} = -\tfrac{1}{c^2}(2E^2)+B^2,
    \end{aligned}
\end{equation}
 where $E_{\mu} \equiv v^{\nu}F_{\mu\nu}$ and $B_{\mu\nu} \equiv h_{\mu}^{\sigma}h_{\nu}^{\rho}F_{\sigma\rho}$.
 Similarly, for the second invariant we get:
 \begin{equation}
{(*F)}_{\mu\nu}F^{\mu\nu} = -\tfrac{1}{c^2}(v^{\mu}v^{\nu}h^{\sigma\rho}+v^{\sigma}v^{\rho}h^{\mu\nu}){(*F)}_{\mu\sigma}F_{\nu\rho}\,.  
 \end{equation}
However, since  ${(*F)}_{\mu\sigma}= g_{\mu\alpha}g_{\sigma\beta}\epsilon^{\alpha\beta \kappa\sigma}F_{\kappa\sigma}$, we recover 
\ba
 {(*F)}_{\mu\nu}F^{\mu\nu} &=& -\tfrac{1}{c^2}(v^{\mu}v^{\nu}h^{\sigma\rho}+v^{\sigma}v^{\rho}h^{\mu\nu})\nonumber\\
 &&\times g_{\mu\alpha}g_{\sigma\beta}\epsilon^{\alpha\beta \kappa \sigma}F_{\kappa\sigma}F_{\nu\rho}\,.       
\ea
By expanding $g_{\mu\nu} = -c^2T_{\mu}T_{\nu}+h_{\mu\nu}$ and performing the contraction, while extracting the $c$ dependence out of $\epsilon^{\alpha\beta \kappa\sigma}$ and using the fact that $v^{\mu}T_{\mu}=-1$, we finally get
\begin{equation}
   \begin{aligned}    {(*F)}_{\mu\nu}F^{\mu\nu} &=\tfrac{1}{c}  (v^{\nu}T_{\alpha}h_{\beta}^{\rho}+v^{\rho}T_{\beta}h_{\alpha}^{\nu})\epsilon^{\alpha\beta \kappa\sigma}F_{\kappa\sigma}F_{\nu\rho} \\&=\tfrac{2}{c} B^{\beta}E_{\beta}\,,
   \end{aligned} 
\end{equation}
where $B^{\beta} \equiv T_{\alpha}\epsilon^{\alpha\beta \mu\nu}F_{\mu\nu}$\,.

\subsection{Electric theory}\label{Maxwell eq}
The electric theory derives from the following Lagrangian:
\ba
\mathcal{L}_{\text{el}} &=& -K^2+K_{\mu\nu}K^{\mu\nu} + 2 (v^{\sigma}\partial_{\sigma}\phi)^2 \nonumber\\
&& +  \tfrac{1}{2}e^{4\phi}(v^{\sigma}\partial_{\sigma}\kappa)^2+ 2 e^{-2\phi}E^2\;.
\ea
Varying the action with respect to $v^{\mu}$ and $h_{\mu\nu}$ yields the constraints \eqref{electric eqns} stated in the main text.

The variation with respect to the dilaton gives the dilatonic equation of motion
\begin{equation}
2v^{\sigma}v^{\rho}\nabla_{\sigma}\partial_{\rho}\phi + e^{4\phi}(v^{\sigma}\partial_{\sigma}\kappa)^2 + 2 e^{-2\phi}E^2=0\,,  
\end{equation}
while the variation with respect to the axion gives the axionic equation:
\begin{equation}
v^{\sigma}v^{\rho}\nabla_{\sigma}\partial_{\rho}\kappa + 4 (v^{\sigma}\partial_{\sigma}\phi)(v^{\rho}\partial_{\rho}\kappa)=0\,.
\end{equation}
Finally, the variation with respect to $A_{\beta}$ gives the following  Maxwell's equations:
\begin{equation}
v^{\alpha}\nabla_{\alpha} (e^{-2\phi}E^{\beta})- v^{\beta}\nabla_{\mu}(e^{-2\phi}E^{\mu})=0\,.
\end{equation}
By contracting with $h_{\beta\mu}$, we get $v^{\alpha}\nabla_{\alpha}(e^{-2\phi}E_{\mu})=0$,
and thence also $\nabla_{\mu}(e^{-2\phi}E^{\mu})=0$. The Maxwell equations thus read
\begin{equation}
v^{\alpha}\nabla_{\alpha} (e^{-2\phi}E^{\beta})=0\,, \quad  \nabla_{\mu}(e^{-2\phi}E^{\mu})=0\,.
\end{equation}

Let us now consider some special cases.
First, let us assume that we have the full EMDA theory, with the dilaton field $\phi$ present (and obeying the corresponding dilatonic equations of motion). Assuming further that  $\phi$ and $\kappa$ are only functions of $r$ and $\theta$, the axionic equation is automatically satisfied, while the dilatonic equation reduces to $E^2=0$. However, since the norm of $E_\mu$ is calculated using the Riemannian metric $h_{\mu\nu}$, this necessarily implies 
\be 
E_\mu=0\,,
\ee 
fulfilling trivially the remaining Maxwell equations.

On the other hand, considering a simpler theory without a dilaton, and $\kappa=\kappa(r,\theta)$, the axionic equations automatically hold, and we recover the  ``standard electrostatics'' equations of motion: 
\begin{equation}
v^{\alpha}\nabla_{\alpha} E^{\beta}=0\,, \quad  \nabla_{\mu}E^{\mu}=0\,.
\end{equation}
In this case, one has a possibility of having a non-trivial electric fields in the stationary and axisymmetric case.

\subsection{Mixed theory}
The mixed theory derives from the following Lagrangian:
\be 
\mathcal{L}_{\text{mix}} =  -2 \kappa E_{\mu}B^{\mu}\,.
\ee  
Since this action is independent of the metric, the Einstein equations are trivial.
The axionic equation for this Lagrangian is given by
\begin{equation}
    E_{\mu}B^{\mu}=0\,,
\end{equation}
while the variation w.r.t the vector potential yields: 
\be
    \nabla_\kappa(\kappa v^\kappa B^\sigma- \kappa v^\sigma B^\kappa + 2 \kappa E_{\mu}T_{\nu}\epsilon^{\mu\nu \kappa\sigma})=0\,.
\ee
Contracting with $h_{\sigma\alpha}$, we get that $\nabla_\kappa(\kappa B^\kappa)=0$. Thence the corresponding vector  equations are:
\be 
\nabla_\kappa(\kappa B^\kappa)=0\,,\quad 
v^\kappa\nabla_\kappa (\kappa B^\alpha)+2\nabla_\kappa(\kappa E_\mu T_\nu \epsilon^{\mu\nu\kappa\alpha})=0\,.
\ee 
Note that for $\kappa=const$ these are just the Bianchi identities and are trivially satisfied (reflecting the fact that in that case of constant axion the action ${\cal L}_{\text{mix}}$ is topological).

\end{document}